\documentclass[twocolumn,showpacs,preprintnumbers,amsmath,amssymb,pr,prc]{revtex4-1}
\usepackage{epsfig}
\usepackage{bbm}
\usepackage{bm}

\begin{document}

\title{Rotational bands in the continuum illustrated by $^{8}$Be results}

\author{E. Garrido$\:^1$, A.S. Jensen$\:^2$, D.V. Fedorov$\:^2$}
\affiliation{$^1$ Instituto de Estructura de la Materia, CSIC,
Serrano 123, E-28006 Madrid, Spain}
\affiliation{$^2$ Department of Physics and Astronomy, Aarhus University, 
DK-8000 Aarhus C, Denmark} 

\date{\today}

\begin{abstract}
We use the two-alpha cluster model to describe the properties of
$^{8}$Be. The rotational energy sequence of the $(0^+,2^+,4^+)$
resonances are reproduced with the complex energy scaling technique
for Ali-Bodmer and Buck-potentials.  However, both static and
transition probabilities are far from the rotational values.  We trace
this observation to the prominent continuum properties of the $2^+$
and $4^+$ resonances.  They resemble free continuum solutions although
still exhibiting strong collective rotational character.  We compare
with cluster models and discuss concepts of rotations in the continuum
in connection with central quantities as transition probabilities,
inelastic cross sections and resonance widths.  We compute the $6^+$
and $8^+$ $S$-matrix poles and discuss properties of this possible
continuation of the band beyond the known $4^+$ state.  Regularization
of diverging quantities are discussed in order to extract observable
continuum properties.  We formulate division of electromagnetic
transition probabilities into interfering contributions from
resonance-resonance, continuum-resonance, resonance-continuum, and
continuum-continuum transitions.
\end{abstract}


\maketitle

\section{Introduction}

 Rotational motion is well defined in classical physics where an inert
structure is rotating as a rigid body around its center of mass. The
two integrals of motion, energy and angular momentum, are continuous
quantities in classical physics. In quantum physics the angular
momentum is always quantized by integer or half integer quantum
numbers, and the energy assumes discrete and continuous values for
bound and unbound states, respectively.  Furthermore, to exhibit
rotational motion quantum systems must have an intrinsic state
deviating from total spherical symmetry \cite{bohrII}.

The signature of rotating quantum systems is a sequence of excited
states with energies following the $J(J+1)$-rule where $J$ is the
angular momentum quantum number.  However, this necessary condition is
not sufficient as the underlying wave functions for different $J$
simultaneously must describe the same rotating structure.  The ratio
of electromagnetic transition probabilities from one of these states
to another are observables given by simple geometric factors depending
only on the angular momentum quantum numbers.  Thus a rotational band
is defined as the sequence of states arising from quantization of the
rotational motion of an (almost) frozen deformed structure
\cite{bohrII}.

The concept of quantum mechanical rotational motion then relies on
discrete quantum states each described by a wave function.  Rotational
states are abundant in molecules and nuclei. 
In molecules numerous rotational states are present \cite{bohrII}. They extent 
to both high angular momenta and relatively high excitation energies. 
The highly excited molecular resonant states in e.g. the $^{24}$Mg nucleus 
represent interesting combinations of molecular structures found in nuclei 
\cite{che10}. These quasimolecular structures were already observed in 
the early 60's in $^{12}$C+$^{12}$C elastic scattering reactions \cite{bro60}.
They are 
in general very well defined even in the continuum where they can decay through 
non-electromagnetic channels, i.e. either by a non-adiabatic molecular process 
or by electron emission through a tunneling process.  The coupling to 
vibrational states is usually responsible for photon emission 
\cite{mcg81,met82,bec09}, as e.g. the $E2$ transitions discussed in the present paper.  
In nuclear physics it has been customary to treat excited states as bound states even when
it is well known that they are embedded in a continuum of states
\cite{wir00,ara04,oer06}.  This includes the many cases where
spontaneous decays are measured \cite{tho04,fre07,alv08,alv10}, and a
width thereby attached to these resonances \cite{til04}. Such
approximations are usually very well justified, first of all in
experimental investigations where pronounced and narrow peaks are
detected.  A measured width is a mixture of intrinsic lifetime,
reaction or decay times and detector resolution, but often the nuclear
states are sufficiently stable to allow population and extraction of
the lifetime.

In theoretical treatments the bound state approximation is very
convenient, since the continuum is much harder to describe.  Most
calculations employ a restrictive basis where the continuum does not
enter, either because it is absent from the start or because it has
been discretized.  In spite of the many successes it is clear that
resonance states do not have a well defined energy, and in principle
they cannot be described by a single wave function.  The difficulties
are increasing with decreasing lifetimes (increasing widths) of these
continuum structures. At some point the widths are so large that the
state has disappeared into the continuum background.  However, much
smaller widths already require clarification of the concept, and in
particular how rotational states can be meaningfully understood.

This basic theme of continuum properties is unavoidable in modern
nuclear physics where far off beta stability and excited states are in
focus \cite{jen04,tho04,pfu12}.  A few years ago a ${\cal
  B}^{(E2)}$-transition ($4^+ \rightarrow 2^+$) was measured in $^8$Be
\cite{dat05} and found to be consistent with previous calculations of
both $\alpha$-$\alpha$ bremsstrahlung cross sections
\cite{lan86,lan86b,gar13} and Greens function Monte Carlo ${\cal
  B}^{(E2)}$-results \cite{wir00}.  However, both measurement and
theories were very far from the rotational prediction and from
comparable classical microscopic cluster model results, see
e.g. \cite{har72,hor70}. This is in spite of the agreement in the rotational
energy sequence. Furthermore, all models agree on the pronounced
deformed $\alpha$-$\alpha$ cluster structure of the $^8$Be-nucleus
\cite{wir00,oer06}.

Thus, even the simplest possible two-body nuclear structure already
presents the problem, which has to be related to the behavior of
continuum structures when the resonances are unmistakenly present but
the widths are not negligibly small. The problem may lie either in the
neglected polarization of the intrinsic $\alpha$-structure for cluster
models, in precise definitions of ${\cal B}^{(E2)}$-values for continuum
models, in a spatially too confining basis in shell models, or in
genuinely unexpected structure of the resonance wave functions.

This paper is based on numerical results for two interacting
alpha-particles. Since an $\alpha$-particle is a composite structure
and we repeatedly use the word clusters and various types of related
models, we shall specify our corresponding definitions.  A cluster is
an entity of particles, which can be anything from one genuine
point-like particle to a group of many correlated particles, which
preferentially effectively act collectively as one particle.  In the
present context the cluster is a group of bound particles with
properties that essentially can be described as one particle but
perhaps with an intrinsic structure.  Specifically we are here only
concerned with the $\alpha$-particle which conceptually in the first intrinsic
layer consists of two neutrons and two protons. The deeper-lying
layers involve virtual mesons, and further on the quark-gluon
intrinsic structures of the nucleons.

Cluster models can then describe structures of (and perhaps reactions
between) clusters where few-body properties are dominating. These
properties may be derived from any of the deeper layers of intrinsic
structures.  We shall in this paper stay with point-like
$\alpha$-clusters, perhaps with finite radius, and each with at most
an underlying layer of four nucleons.  The Pauli principle is
accounted for by an effective $\alpha-\alpha$ interaction.  A
classical cluster model is then naturally defined as a model for
point-like interacting particles without any intrinsic degrees of
freedom, and with an effective phenomenological interaction.  In
microscopic cluster models the effective interaction is in principle
derived from an approximation to the nucleon-nucleon interaction and
the Pauli principle.  A classical microscopic cluster model is a
microscopic version with an old relatively simple nucleon-nucleon
interaction.

The purpose of this paper is to clarify the concepts of rotational
states in the continuum. Taking the case of $^8$Be as an example, we
shall pin-point the problems, clarify the definitions, show how to
avoid pit-falls, and give the minimum requirements for future model
computations of continuum properties.  We first in section II briefly
describe the basic ingredients, the pertinent formalism, notation and
definitions. Then in section III we discuss the calculated numerical
results in connection with the rotational model, that is energies and
transition probabilities.  In section IV we discuss various features
of the transition matrix elements, validity conditions for appearance
of collective rotations, and rotational states in heavier nuclei.  In
section V we finally give a summary and the conclusions.

\section{The basic ingredients}

The rotational energy sequence is defined by \cite{bohrII}
\begin{eqnarray}
E_{\ell} = E_0 + \frac{\hbar^2\ell(\ell+1)}{2 {\cal I}} \;,   \label{eq1a}
\end{eqnarray}
where $\ell$ is the angular momentum and ${\cal I}$ is the moment of
inertia around the rotation axis.  $E_0$ is the energy of the lowest
state, $\ell=0$, in the rotational band.

We shall aim at computing the decay probability, which is simply
related to the transition strength ${\cal B}^{(E\lambda)}(\ell
\rightarrow \ell^\prime)$. We shall now only consider $\lambda = 2$
with the intended application on a system of two $\alpha$-particles
described in the relative coordinate system.  The $\lambda = 2$
electric multipole transition is the lowest transition possible and
the contribution from $\lambda = 4$ is orders of magnitude
smaller. Generalization of the formalism to $\lambda$-values different
from $2$ is straightforward, see \cite{gar13}.

The immediate theoretical problem is that ${\cal B}^{(E2)}$ for
continuum transitions is not uniquely defined. It is necessary to
start with quantum mechanical observables, and from these define
meaningful quantities to describe the desired decay probabilities. One
unavoidable requirement is that relations between observables and
derived quantities must be identical to the established expressions in
the limit of bound states and very narrow resonances.

\subsection{Cross sections}

We begin with the differential cross section for emission of a photon
of energy $E_{\gamma}$, from an initial two-body continuum state of energy $E$,
arriving at a final continuum state of energy $E^{\prime}$.  The
differential cross section for this process is given by
\cite{lan86b,gar13}:
\begin{eqnarray}
\frac{d\sigma^{(E2)}}{dE_\gamma} \bigg|_{\ell \rightarrow \ell^\prime}=
\frac{\pi^2 Z^2e^2}{ 15 k^2}
(2\ell+1)
\left(\frac{E_\gamma}{\hbar c} \right)^{5} \nonumber  \\ 
 \times \bigg| \langle \ell 0; 2 0| \ell^\prime 0\rangle
\int_0^\infty u_\ell(E,r) r^{2} u_{\ell^\prime}(E^\prime,r) dr
 \bigg|^2 \;, 
\label{eq3} 
\end{eqnarray}
where $Z=2$ for two $\alpha$-particles, $e^2=1.4400$~MeV~fm,
$E_\gamma=E-E^\prime$ is the energy of the emitted photon, $\ell$ and
$\ell^\prime$ are the relative angular momenta between the two
particles in the initial and final state, and $k^2=2 \mu E/\hbar^2$
($\mu$ is the reduced mass of the two-body system).

The radial wave functions $u_\ell$ and $u_{\ell^\prime}$ describe two-body continuum
structures, and they are solutions of the radial two-body Schr\"{o}dinger equation 
for the initial and final states, respectively. They obey the large-distance boundary condition
\begin{equation}
u_\ell(E,r) \stackrel{r \rightarrow \infty}{\longrightarrow} 
 \sqrt{\frac{2 \mu}{\pi \hbar^2 k}}
  \left[ \cos\delta_\ell F_\ell(kr) + \sin\delta_\ell G_\ell(kr)\right],
\label{asymp}
\end{equation}
where $F_\ell$ and $G_\ell$ are the regular and irregular Coulomb
functions, $\delta_\ell$ is the nuclear phase shift, and the normalization
constant is determined by the orthogonality condition:
\begin{equation}
\int_0^\infty u_\ell(E,r) u_\ell(E^\prime,r) dr=\delta(E-E^\prime).
\label{ener}
\end{equation}

A delicate point in the calculation of the cross section refers to the procedure employed to
obtain the integral in Eq.(\ref{eq3}). The continuum wave functions do not drop off at 
infinity, and the radial integrals oscillate with larger and larger amplitudes as $r$ increases.
This presents a severe numerical challenge. To overcome this problem  we shall in  this work
employ the Zel'dovich prescription \cite{zel60}, which
introduces the regularization factor, $e^{-\eta^2 r^2}$, in the
radial integrand. This eliminates the many large amplitude oscillations at large distances
which in any case mathematically can be shown to cancel out.
The correct result is then obtained in the limit of zero value for the Zel'dovich parameter $\eta$.
Fortunately, this method removes the unwanted large-distance oscillations and the
remaining physical results are uniquely defined, since they are stable
for sufficiently small values of $\eta$.  A formal discussion of this kind of integrals can be found 
in \cite{gya71,gya72}.

The total cross section for emitting a photon of any energy, possibly
confined to a pre-decided final energy interval $\Delta E'$, is
obtained by integration
\begin{eqnarray}
 \sigma^{(E2)}_{\ell \rightarrow \ell^\prime}(E)=\int_{\Delta E'} \left. 
 \frac{d\sigma^{(E2)}}{dE_\gamma}\right|_{\ell \rightarrow \ell^\prime} 
 \hspace*{-5mm}(E) \;dE_\gamma  \;,
\label{eq4a}
\end{eqnarray}
where we implicitly assume that $E'=E-E_{\gamma}$.  The confining
interval can be decided in a practical experimental measurement, for
example as a window around a resonance energy in the final state. This selects then
approximately resonance properties without continuum admixtures,
although this in practice easily becomes ambiguous at the desired level
of accuracy.

In the case of a transition into a bound state ($u_{\ell'}$ describing
a bound state with a well defined final energy), Eq.(\ref{eq3}) is
still valid, with the only difference that the r.h.s. of the equation
already gives the total cross section $\sigma^{(E2)}_{\ell \rightarrow
  \ell^\prime}(E)$ instead of the differential one (note that the
different dimension of a bound state wave function compared to the one
of a continuum wave function, which can be seen for instance from
Eq.(\ref{asymp}), makes the change dimensionally consistent).  For
this particular case of transition into a bound state the total
cross section and the strength function are related by the well known
expression
\begin{equation}
 \sigma^{(E2)}_{\ell \rightarrow \ell^\prime}(E)=\frac{2(2\pi)^3}{75}
 \frac{1}{k^2} \left( \frac{E_\gamma}{\hbar c} \right)^{5}  
(2\ell+1)
 \frac{d{\cal B}^{(E2)}}{dE}(\ell\rightarrow \ell^\prime),
\label{eq1}
\end{equation}
which can be easily generalized to the case of transitions between continuum states as:
\begin{equation}
\frac{d\sigma^{(E2)}}{dE_\gamma} \bigg|_{\ell \rightarrow \ell^\prime}=
 \frac{2(2\pi)^3}{75} \frac{1}{k^2} \left( \frac{E_\gamma}{\hbar c} \right)^{5}  
(2\ell+1)
 \frac{d{\cal B}^{(E2)}}{dEdE'}(\ell\rightarrow \ell^\prime).
\label{eq7}
\end{equation} 

From Eqs.(\ref{eq3}) and (\ref{eq7}) we can easily identify:
\begin{eqnarray}
\lefteqn{ \hspace*{-0.7cm}
 \frac{d{\cal B}^{(E2)}}{dEdE'}(\ell\rightarrow \ell^\prime) = } \nonumber \\ & &
\frac{5e^2}{4\pi}
\langle \ell 0; 2 0| \ell^\prime 0\rangle^2
\bigg|
\int_0^\infty u_\ell(E,r) r^{2} u_{\ell^\prime}(E^\prime,r) dr
 \bigg|^2 \;,
\label{dbde}
\end{eqnarray}
which agrees with the standard definition:
\begin{eqnarray}
\lefteqn{ \hspace*{-0.7cm}
 \frac{d{\cal B}^{(E2)}}{dEdE'}(\ell\rightarrow \ell^\prime) = } \nonumber \\ & &
\sum_{\mu,m_{\ell'}}
\left|
\langle \Psi_{\ell',m_{\ell'}}(E',\bm{r}) | e r^2 Y_{2,\mu}(\Omega_r)| \Psi_{\ell,m_{\ell}}(E,\bm{r} \rangle
\right|^2\;,
\end{eqnarray}
where now $\Psi_{\ell,m_{\ell}}(E,\bm{r})=u_\ell(E,r) Y_{\ell,m_\ell}(\Omega_r)/r$ is the full initial 
two-body wave function (and similarly for the final state wave function $\Psi_{\ell',m_{\ell'}}(E',\bm{r})$).

In practical continuum calculations it is rather frequent to employ
some kind of discretization procedure. In this way the continuum is
described by a set of states with discrete energies $\{E_i\}$, whose
corresponding radial wave functions $\{u_{\ell}^{(i)}(E_i,r)\}$
usually are normalized following the standard bound-state rule:
\begin{equation}
\int_0^\infty u_\ell^{(i)}(E_i,r) u_\ell^{(j)}(E_j,r) dr=\delta_{ij}.
\label{ort}
\end{equation}

Making use of the relation between the Dirac and Kronecker deltas ($\delta_{ij}=
\lim_{\Delta E \rightarrow 0} \Delta E \;\delta(E_i-E_j)$, with $\Delta E$ the energy separation between the 
two states) and the continuum normalization rule, Eq.(\ref{ener}), it is possible to relate the
continuum ($u_\ell$) and the discretized continuum ($u_\ell^{(i)}$) wave functions by:
\begin{equation}
u_\ell^{(i)}(E_i,r)=\lim_{\Delta E \rightarrow 0} \sqrt{\Delta E} \; u_\ell(E_i,r),
\end{equation}
from which we have 
\begin{equation}
\langle u_\ell(E,r) | u_\ell^{(i)}(E_i,r)\rangle =
\lim_{\Delta E \rightarrow 0} \sqrt{\Delta E} \; \delta(E-E_i).
\end{equation}

Finally, the expression above and the closure relation 
$\mathbbm{1}=\sum_i |u_\ell^{(i)}(E_i,r)\rangle \langle u_\ell^{(i)}(E_i,r) |$
lead to:
\begin{eqnarray}
\lefteqn{
\left|
\int_0^\infty u_\ell(E,r) r^2 u_{\ell^\prime}(E^\prime,r) dr
\right|^2= }  \label{eq12}\\ &&
 \sum_{i,j} \delta(E-E_i) \delta(E^\prime-E^\prime_j) 
\left|
\int_0^\infty u_\ell^{(i)}(E_i,r) r^2 u_{\ell'}^{(j)}(E^\prime_j,r) dr
\right|^2,  \nonumber
\end{eqnarray}
which implies that after discretization of the continuum, the
differential cross section Eq.(\ref{eq3}) should be computed with the
replacement indicated in Eq.(\ref{eq12}), where $i$ and $j$ run over
the discrete initial and final states, respectively.

Thanks to the delta functions, the integral Eq.(\ref{eq4a}) can be
trivially calculated, and we get for the integrated cross section:
\begin{eqnarray}
\sigma^{(E2)}_{\ell \rightarrow \ell^\prime}(E) &=& \frac{4\pi^2 e^2}{15 k^2} (2\ell+1)  
\langle \ell 0; 2 0| \ell^\prime 0\rangle^2  \label{eq14} \\ & & \hspace*{-2cm}
\times \sum_{i,j} \left(\frac{E_\gamma}{\hbar c} \right)^5 \delta(E-E_i) 
\left|
\int_0^\infty u_\ell^{(i)}(E_i,r) r^2 u_{\ell'}^{(j)}(E^\prime_j,r)dr 
\right|^2 . \nonumber
\end{eqnarray}
In the same way that the integration Eq.(\ref{eq4a}) can be restricted to final energies within
some chosen final energy window, in Eq.(\ref{eq14}) the summation over $j$ can also be restricted to those
discrete final states whose energy $E'_j$ is contained in the chosen energy window. However, in order
to reach a sufficient accuracy in the calculation, it is necessary to have a significant amount of
discrete final energies within that window.

From Eqs.(\ref{eq12}) and (\ref{dbde}) it is also evident that after discretization of the continuum the
differential transition strength takes the form:
\begin{eqnarray}
&&\frac{d{\cal B}^{(E2)}}{dE}(\ell\rightarrow \ell^\prime) = 
 \frac{5e^2}{4\pi} \langle \ell 0; 2 0| \ell^\prime 0\rangle^2 
\label{eq10}\\ & &
\times \sum_{i,j} \delta(E-E_i) 
\left|
\int_0^\infty u_\ell^{(i)}(E_i,r) r^2 u_{\ell'}^{(j)}(E^\prime_j,r)dr 
\right|^2 , \nonumber
\end{eqnarray}
from which we can in principle integrate over $E$ and obtain the total transition strength:
\begin{eqnarray}
&& {\cal B}^{(E2)}(\ell\rightarrow \ell^\prime) = 
 \frac{5e^2}{4\pi} \langle \ell 0; 2 0| \ell^\prime 0\rangle^2 
\label{eq13a}\\ & &
\times \sum_{i,j}  
\left|
\int_0^\infty u_\ell^{(i)}(E_i,r) r^2 u_{\ell'}^{(j)}(E^\prime_j,r)dr 
\right|^2 , \nonumber
\end{eqnarray}
where again the summation over $j$ could be restricted to the chosen final energy window. 

For a given transition the total transition strength and the decay probability, $\Gamma_{\gamma}$, 
are related through
\begin{eqnarray}
\Gamma_{\gamma}^{(\ell\rightarrow \ell^\prime)} = \frac{4\pi}{75}
\left( \frac{E_\gamma}{\hbar c} \right)^{5} 
{\cal B}^{(E2)}(\ell\rightarrow \ell^\prime)  \;.
\label{eq2} 
\end{eqnarray}

\subsection{Structure extraction as ${\cal B}^{(E2)}$-values}
\label{secbe2}

In direct calculations ${\cal B}^{(E2)}$-values could in principle be
obtained by use of expressions like Eq.(\ref{eq10}) or (\ref{eq13a}).
However, an indiscriminate sum over initial and final states makes the
result rather meaningless. The information about resonance properties
is completely washed out and, even worse, weighted at the wrong
energies.  Furthermore, due to the undesired divergence produced by the
soft-photon contribution ($E^\prime \rightarrow E$ or 
$E^\prime_j \rightarrow E_i$) \cite{gar13}, the calculation itself is pretty
complicated.

Instead, it is necessary to return to the observable cross sections, and 
then extract the transition strength from expressions like Eqs.(\ref{eq7}) 
and (\ref{eq14}).  We have especially investigated two rather
different methods to obtain ${\cal B}^{(E2)}$ values (see ref.\cite{gar13}).  
The first assumes a Breit-Wigner shape of the cross section (\ref{eq4a}) or (\ref{eq14}) 
around the energy of the resonance in the initial channel.
The resonance width is energy dependent, but at the resonance energy it has to be equal to the 
bare width of the resonance.  The matching to the computed cross section provides the decay
probability, $\Gamma_{\gamma}$, which through Eq.(\ref{eq2})
immediately gives ${\cal B}^{(E2)}$.

This method fundamentally assumes a Breit-Wigner shape of the cross
section.  This is only correct in a rather narrow range of energies
around the resonance.  This apparent restriction is perhaps physically
reasonable since it corresponds to transitions between resonance
peaks.  The possibly undesired background continuum contributions are
then eliminated  (compare to the discussion in Section IV.A).

The second method employs Eq.(\ref{eq7}) by integrating over the
initial and final energies, $E$ and $E'$, which run over the chosen 
initial and final energy windows.  If the photon energy, $E_{\gamma}$, were constant,
this would immediately provide a ${\cal B}^{(E2)}$-value.  However,
since this assumption is incorrect, we must use an average value of
$E_{\gamma}^5$ in order to extract ${\cal B}^{(E2)}$.  Thus, we
define
\begin{eqnarray}
{\cal B}^{(E2)} \propto  \frac{\int \sigma^{(E2)}(E) dE }{<E_{\gamma}^5>} \;,
\label{eq9} 
\end{eqnarray}
where $E_{\gamma}$ is chosen as the difference between the energy of
the cross section peak position and the energy of the resonance in the
final state.  Again, this assumes information about resonance
positions but as with the first method (some of) the continuum
background contributions are eliminated.  Unavoidably the sensitivity
is noticeable to rather small variations around a chosen $E_{\gamma}$
due to the power of $5$ for ${\cal B}^{(E2)}$ transitions (see \cite{gar13}
for details).

These quantities are easily defined and measured for reactions where
bound states are involved, and model calculations are numerous
\cite{bohrII}.  This bound state limit is perfectly correct and well
defined.  However, difficulties begin to pile up when members of the
rotational band reach into the continuum and acquire a width for decay
through channels that lead to states outside the band.  The energies
may be relatively simply measured, and analyzed as peaks with widths
populated in reactions or perhaps in decays from other channels
\cite{pfu12}.  Theoretical techniques to deal with these problems are
discussed in details in Ref.\cite{gar13a}.

Calculations of energies and gamma-widths are ambiguous in the
continuum, especially when the total width is large.  An efficient
method that permits to extract the energy and width of resonances is
the complex scaling method \cite{ho83,moi98}, which after rotation of
the radial coordinates into the complex plane makes the resonances
appear formally as bound states with complex energy.  The resonances
obtained in this way correspond to the poles of the $S$-matrix, and
the real and imaginary parts of the complex energy describe
respectively, the resonance position and half the width. The
corresponding complex rotated resonance wave function is also obtained
by this method.

It may then be illuminating to compare the ${\cal B}^{(E2)}$-value
with those obtained by the precise resonance definition in the complex
scaling method.  Here the resonance wave functions are well defined
and their (complex) energy difference as well. This provides uniquely
a resonance-to-resonance, complex scaled ${\cal B}^{(E2)}$-value.  In
principle, the full transition strength Eq.(\ref{eq13a}) can also be
computed within the complex scaling framework \cite{myo98}. However,
although this method allows a clean and precise extraction of the
resonances, the contact with the observable quantities is sometimes
less direct. In particular, when transitions involve continuum states,
all the difficulties arising from the description of the continuum
states themselves would mix with the interpretation of the complex
scaled transition strength, which furthermore is a complex quantity.

\section{Properties of the $^8$Be states}

We have chosen to illustrate our understanding of continuum structures
with $^8$Be, which is the simplest non-trivial two-body cluster
nucleus. The effective interaction between the two $\alpha$-particles
is very well known from scattering experiments and the subsequent
analysis in terms of partial wave phase shifts. A number of potentials
reproducing the low-energy elastic scattering cross sections are
available. We shall employ the Buck-potential \cite{buc77} and 
version {\it d} of the Ali-Bodmer potentials given in
\cite{ali66}. The Buck potential has two spurious deep-lying
$\alpha-\alpha$ bound states for $s$-waves, and one more for
$p$-waves.  When necessary, the spurious states can be removed by the
construction of a phase equivalent potential \cite{gar99}.

The dependence of the transition probability on the potential is of
basic interest, since the contributing parts of the wave functions are
expected to be at short distances.  Identical phase shifts reflect
identical large-distance properties but with different nodes at short
distances. Thus, interaction dependent transition probabilities may
arise.  In other words the ${\cal B}^{(E2)}$-values could be able to
distinguish between potentials of different short-distance properties,
that is, in particular, between Buck and Ali-Bodmer potentials.
However, as shown in \cite{gar13}, the ${\cal B}^{(E2)}$ transition
strength shows very minor changes when switching from one of the interactions
to the other.

\subsection{Energies and radii}

\begin{figure}
\epsfig{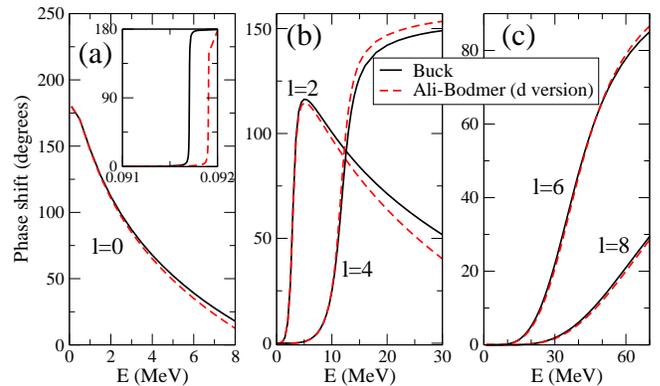}
\caption{(Color online) Phase shifts (in degrees) as a function of the
$\alpha-\alpha$ relative energy obtained with the Buck potential (solid curves) 
and the {\it d}-version of the Ali-Bodmer potential (dashed curves). The 
results for $\ell=0$ (panel a), $\ell=2,4$ (panel b), and $\ell=6,8$ (panel c) 
are shown. The inset in panel a shows the phase shifts in vicinity of the
0$^+$ resonance in $^8$Be.} 
\label{fig1}
\end{figure}

The bosonic nature of the $\alpha$-particle constrains the possible
excited states to have even angular momenta and positive parity.  In
Fig.\ref{fig1} we show the phase shifts as a function of the
$\alpha-\alpha$ relative energy for the $\ell=0,2,4,6,8$ partial
waves. The solid and dashed curves correspond to the results obtained
with the Buck potential and the Ali-Bodmer potential,
respectively. The inset in panel (a) gives the extremely rapidly
varying $s$-wave phase shifts in the vicinity of the $0^+$ resonance
energy in $^8$Be. The computed phase shifts are remarkably similar
for both potentials, even for large relative angular momenta. The
computed $^8$Be spectrum is then expected not to change very much from
one potential to the other.

As already mentioned, the complex scaling method permits an easy
evaluation of the resonance energies and widths. The results obtained
for $^8$Be are given in Table~\ref{tab1} along with the experimentally
known resonance energies and widths \cite{til04}.  The Buck and
Ali-Bodmer potentials both, as expected from the phase shifts, provide
very similar spectra.  Together with the $0^+$, $2^+$, and $4^+$
states whose energies and widths reproduce rather well the
experimental values, both potentials predict a $6^+$ and an $8^+$
state at about 34 MeV and 52 MeV, respectively. No experimental
evidence of these states is known. In any case, the computed widths
for these two last resonances are comparable to their energies and
therefore they can not be considered as well defined resonances.  A
similar proportion between width and energy as for the 2$^+$ and $4^+$
states  would for the same energies give widths of about 2.5 times
smaller values, that is 13 MeV and 21 MeV for the $6^+$ and $8^+$
states, respectively.

An important point to take into account is the fact that the energy of
the $0^+$ resonance is very sensitive to the $\hbar^2/m_\alpha$-value
(with $m_\alpha$ being the mass of the alpha-particle) used in the
calculation.  The experimentally known value is
$\hbar^2/m_\alpha=10.446$ MeV$\cdot$fm$^2$.  This is used for all the
calculations with the Ali-Bodmer potential.  However, when the same
value is used with the Buck potential \cite{buc77} the $0^+$ resonance
appears at 0.18 MeV, almost a factor of 2 higher than the experimental
value.  The potential parameters given in \cite{buc77} are therefore
probably obtained with $\hbar^2/m_\alpha=10.368$ MeV$\cdot$fm$^2$,
which places the $0^+$ resonance at the correct value.  The effects on
the $0^+$ resonance and the $s$-wave phase shifts are given in
table~\ref{tab1} and the solid line in the inset of Fig.\ref{fig1}a.
The other resonances (with $\ell >0$) are much less sensitive to this
change in $\hbar^2/m_\alpha$.  This original value is maintained when
the Buck-potential is used in this paper.

\begin{figure}
\epsfig{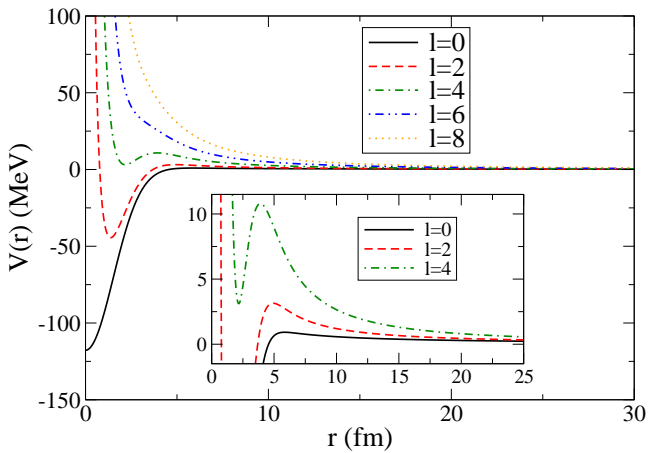}
\caption{(Color online) Total two-body potential (Buck potential +
  Coulomb repulsion + Centrifugal barrier) for the lowest angular
  momenta. The inset shows the details of the potential barrier for
  $\ell=$ 0, 2, and 4. }
\label{fig2}
\end{figure}

The two-body potentials giving rise to all these resonances
are shown in Fig.\ref{fig2}. Together with the nuclear interaction (that
for the figure has been chosen to be the Buck potential) the potentials
shown in the figure contain as well the Coulomb repulsion and the centrifugal
barrier.  For $\ell=0$ the potential barrier is hardly noticeable but the resonance
energy is still smaller and the state experiences an
extremely thick barrier leading to almost bound state properties. For
$\ell=2,4$ the barrier is much higher and thinner but the energy is
not far from the top and the resulting widths are rather large. The
details of the different potential barriers are shown in the inset. For
$\ell=6,8$, the potentials are repulsive. It is
therefore surprising that the $S$-matrix poles apparently are well
defined and independent of the interactions, determined solely from the
phase shifts of the partial waves of smaller $\ell$-values.  For this
reason we include the results for $\ell=6,8$, although a resonance
description is a stretch of this concept. 

\begin{table*}
\begin{center}
\caption{Properties of the five lowest computed resonances in
  $^8$Be. The first two rows give, when available, the corresponding
  experimental energies, $E_r$, and widths, $\Gamma_r$, taken from
  Ref.\cite{til04}. The computed values with the Buck and Ali-Bodmer
  potentials are given by the third and fourth rows, and by the fifth
  and sixth rows, respectively.  All the energies and widths are given
  in MeV. The following four rows give, also for the two
  $\alpha-\alpha$ potentials, the real and imaginary parts of
  $\sqrt{<r^2>}$, computed with the complex scaling method. These
  values are given in fm.  The rows marked $E_r^{(0)}$, $E_r^{(1)}$,
  and $E_r^{(Z_0)}$ are rotational energies (in MeV) defined through
  Eq.(\ref{eq1a}) and the corresponding moments of inertia
  $\hbar^2/(2{\cal I})$, which are denoted by $B_0$ (in MeV) when
  obtained by fitting the energies through Eq.(\ref{eq1a}), $B_1$ when
  obtained from Eq.(\ref{eq4}) with constant $\alpha$-$\alpha$
  distance ($Z_0=3.0$~fm), and $B_{Z_0}$ when obtained from
  Eq.(\ref{eq4}) with angular momentum dependent $\alpha$-$\alpha$
  distance.  While $B_1$ takes the value of 0.621 MeV, $B_0$ and
  $B_{Z_0}$ are angular momentum dependent and they are given in the
  table for each resonance (the values given for $E_r^{(0)}$ have been obtained
  with $B_0=0.475$ MeV, see text).  The last row gives the
  excitation energies obtained in the microscopic cluster model
  \cite{har72}. }
\label{tab1}
\begin{ruledtabular}
\begin{tabular}{|c|ccccc|}
\hline 
 $J^+$  &  $0^+$ & $2^+$ & $4^+$ &  $6^+$ & $8^+$ \\
\hline
$E_r$ (Exp.)        &         0.0918          & $2.94\pm0.01$   &   $11.35\pm0.15$      &  ---   &  ---      \\
$\Gamma_r$  (Exp.)  & $(5.57\pm0.25) 10^{-6}$ & $1.51\pm0.02$   &     $\sim 3.5  $      &  ---   &  ---      \\
\hline
 $E_r$ (Buck)       &          0.091           &      2.88       &      11.78            &  33.55  &  51.56   \\
 $\Gamma_r$ (Buck)  &   $3.6\cdot10^{-5}$      &      1.24       &       3.57            &  37.38  &  92.38   \\
\hline
 $E_r$  (Ali-Bodmer d)   &          0.092           &      2.90       &      11.70            &  34.38  &  53.65   \\
$\Gamma_r$ (Ali-Bodmer d)&   $3.1\cdot10^{-6}$      &      1.27       &       3.07            &  37.19  &  93.74   \\
\hline
Re{$\sqrt{<r^2>}$} (Buck) & 5.61 & 3.51 & 2.93 & 2.82 & 2.76 \\
Im{$\sqrt{<r^2>}$} (Buck) & 0.01 & 1.29 & 0.82 & 1.44 & 1.77 \\
\hline
Re{$\sqrt{<r^2>}$} (Ali-Bodmer d) & 5.80  & 3.58 & 2.91 & 2.70 & 2.73 \\
Im{$\sqrt{<r^2>}$} (Ali-Bodmer d) & 0.001 & 1.24 & 0.76 & 1.40 & 1.73 \\
\hline
 $E_r^{(0)} = E_0+ B_0 J(J+1)$  & $E_0=0.091$ & 2.9  &  9.6   & 20.0  & 34.3   \\
\hline
 $B_0$ & --- & 0.475  &  0.563   & 0.807  & 0.721  \\
\hline
 $E_r^{(1)} = E_0+ B_1 J(J+1)$   & $E_0=0.091$ & 3.8  & 12.5   & 26.2  & 44.8   \\
\hline
 $E_r^{(Z_0)} = E_0+ B_{Z_0} J(J+1)$  & $E_0=0.091$ & 3.2  &  12.9   & 28.5  & 49.1   \\
\hline
 $B_{Z_0}$  & --- & 0.511  &  0.639   & 0.677  & 0.680   \\
\hline
 $E_r$ \cite{har72}  & --- & 3.8  &  13.5   & 30.5  & 49.7   \\
\hline
\end{tabular}
\end{ruledtabular}
\end{center}
\end{table*}

With the complex rotated wave functions of resonances at hand it is
possible to compute the corresponding expectation values of $r^2$,
which for resonances are complex numbers in contrast to the real
values obtained for bound states even if the corresponding wave 
functions have been complex rotated.  As
discussed in \cite{moi98}, the real part of the expectation value of a
given complex rotated operator can be understood as a corresponding
average value over continuum wave functions in a range of energies
around the resonance.  It is then tempting to associate the imaginary
part with an uncertainty of the same expectation value.  This is
analogous to the energy associated with the expectation value of the
complex rotated hamiltonian.

In Table~\ref{tab1} the real and imaginary parts of $\langle r^2
\rangle^{1/2}$ for the five resonances found in $^8$Be with the two
different $\alpha-\alpha$ potentials are shown.  Again, both
potentials give very similar values. 
We refer to the imaginary parts as the uncertainty which according to \cite{gya72,moi98}
arises from two sources, i.e. the finite width of the state and the fact 
that the resonance wave function is not an eigenfunction of the $r^2$ operator. 
For the $0^+$ case the
uncertainty in $\langle r^2 \rangle^{1/2}$ is very small, as it has to
be for such a narrow resonance. For the $2^+$ and $4^+$ states the
uncertainty in the size of the resonance is about three or four times
smaller than their average values, while for the $6^+$ and $8^+$ cases
the uncertainty increases up to half the average value.  The computed
real parts of the average values of $\langle r^2 \rangle^{1/2}$ are
quite similar for the $4^+$, $6^+$, and $8^+$ resonances.  By
increasing the relative orbital angular momentum, the two
$\alpha$-particles appear more and more spatially confined, as
discussed below in more details.  This apparent confinement is in
spite of the broad resonance structures arising from being in the
continuum, where extended spatial extension intuitively is expected.
It is worth emphasizing that the radii discussed so far are well
defined theoretical quantities but they are not observables.  The
expectation value of the operator $r^2$ in a continuum wave function
is infinitely large.

The three lowest resonance energies have traditionally been
interpreted as energies of a rotational band following the behavior in
Eq.(\ref{eq1a}) with a structure of two $\alpha$-clusters at a given
distance from each other. Therefore, the value of $B_0=\hbar^2/(2{\cal
  I})$ should be constant, and it could be extracted for instance from
Eq.(\ref{eq1a}) and the energy of the $2^+$ resonance.  This gives a
value of $B_0= 0.475$ MeV, and the resulting energies of the different
levels in the rotational band become those denoted by $E_r^{(0)}$ in
table~\ref{tab1}.  In this case the energy of the $4^+$ resonance
appears at about 2 MeV below the measured (and computed) value, and
the energies of the $6^+$ and $8^+$ states are clearly smaller than
the computed ones. Obviously, the reason is that the $B_0$ value
changes quite a lot when extracted by use of the different resonance
energies, as we can see in table~\ref{tab1}.  The fact that $B_0$ is
clearly energy dependent suggests that the structure of $^8$Be does
not really correspond to a rigid rotor system, i.e., it does not match
with an almost frozen deformed structure.

In any case, if we still assume
that the two $\alpha$-particles are fixed at positions $z=\pm Z_0/2$ on the $z$-axis,
the rigid moment of inertia ${\cal I}_{rig}$ around the $y$-axis is then
given by:
\begin{eqnarray}
{\cal I}_{rig} = \frac{1}{2} m_{\alpha}  Z_0^2  + \frac{4}{5} m_{\alpha} R_{\alpha}^2,  
\label{eq4}
\end{eqnarray}
where $m_{\alpha}$ is the mass of the alpha particle, $\langle
r_\alpha^2\rangle^{1/2}\approx 1.7$ fm is its root mean square radius,
and the corresponding sharp cut-off radius is $R_{\alpha}^2 = 5/3
\langle r_\alpha^2 \rangle$. If we take a $Z_0$ value of 3.0 fm 
we then get $B_1=\hbar^2/(2{\cal I}_{rig})=0.621$ MeV, which is a kind
of average of the four $B_0$ values previously obtained. The estimates of
the resonance energies from Eqs.(\ref{eq1a}) and (\ref{eq4}) are given in 
Table~\ref{tab1} as $E_r^{(1)}$. The agreement of this rotational sequence 
with the experimental and computed values is not perfect but perhaps acceptable.

The fact that the intrinsic structure of $^8$Be is changing with
angular momentum is also evident from the root mean square radii of
the resonances.  In other words, the moment of inertia is also angular
momentum dependent, and it can be obtained for each resonance from
Eq.(\ref{eq4}) by simply assuming that for each of them the two
$\alpha$-particles are located at the corresponding distances,
$Z_0=\mbox{Re}\sqrt{\langle r^2 \rangle}$.  In this way, we get the
$\hbar^2/(2{\cal I}_{rig})$ values given in table~\ref{tab1} as
$B_{Z_0}$, and the energy sequence, Eq.(\ref{eq1a}), given by
$E_r^{(Z_0)}$.  They match very nicely with the energies obtained
directly by solving the two-body problem with the corresponding
$\alpha-\alpha$ potential.  The results for the excitation energies
from a classical microscopic cluster model \cite{har72} are given in the
last row of the table, and they are remarkably similar although
obtained with a completely different two-body nucleon-nucleon
potential.

This latter result seems to confirm that the $^8$Be spectrum has a
rotational character.  However, the best agreement has been obtained
by using different values of $\mbox{Re}\sqrt{\langle r^2 \rangle}$ for
$Z_0$ for each resonance.  The moments of inertia are correspondingly
very different, and surprisingly the largest variation in
$\sqrt{\langle r^2 \rangle}$ is found for the lowest two, 0$^+$ and
2$^+$, of the three experimentally known states.  This fact reveals
that the idea of $^8$Be as a rigid rotor of two alpha particles
separated by a given distance is questionable.  Only for the 4$^+$,
$6^+$, and $8^+$ states the distance remains roughly the same, and
therefore also ${\cal I}_{rig}$ is more stable.  However, it is worth
emphasizing that the intrinsic $\alpha$-particle structure still is
maintained, although the particles are located at different separation.

\subsection{Transitions}

In Section \ref{secbe2} we have described two different methods to obtain
the transition strength ${\cal B}^{(E2)}$ from the computed cross section
for a given transition.  The first one assumes a Breit-Wigner shape for the
cross section in the vicinity of the resonance energy for the incident channel.
This fact permits to extract the decay probability $\Gamma_\gamma$ for that
transition and, from Eq.(\ref{eq2}), obtain then the transition strength. We shall
denote the strength computed in this way as ${\cal B}^{(E2)}_\gamma$.

\begin{figure}
\epsfig{file=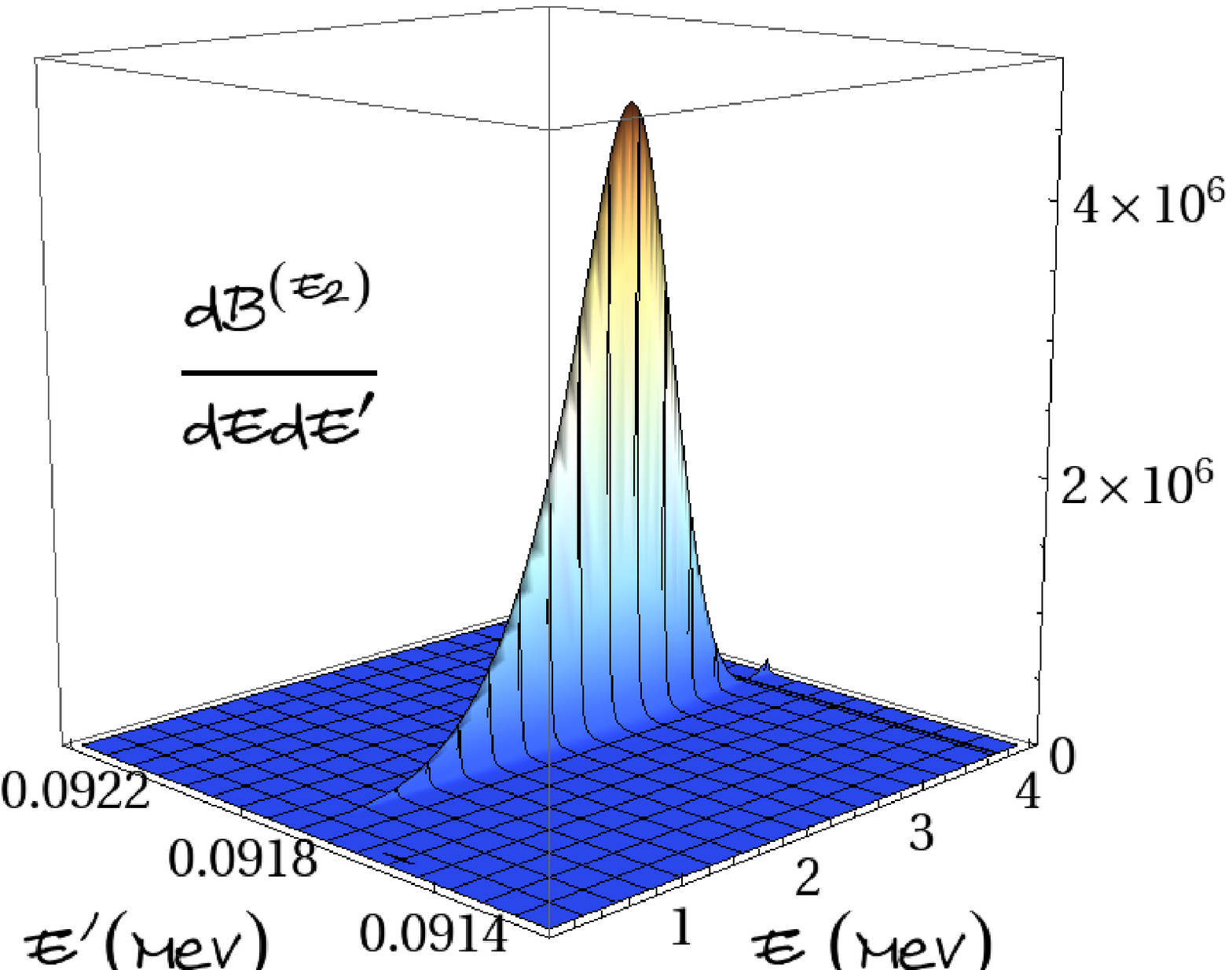,width=8.5cm,angle=0}
\vspace*{0cm} \hspace*{-1.5cm}
\epsfig{file=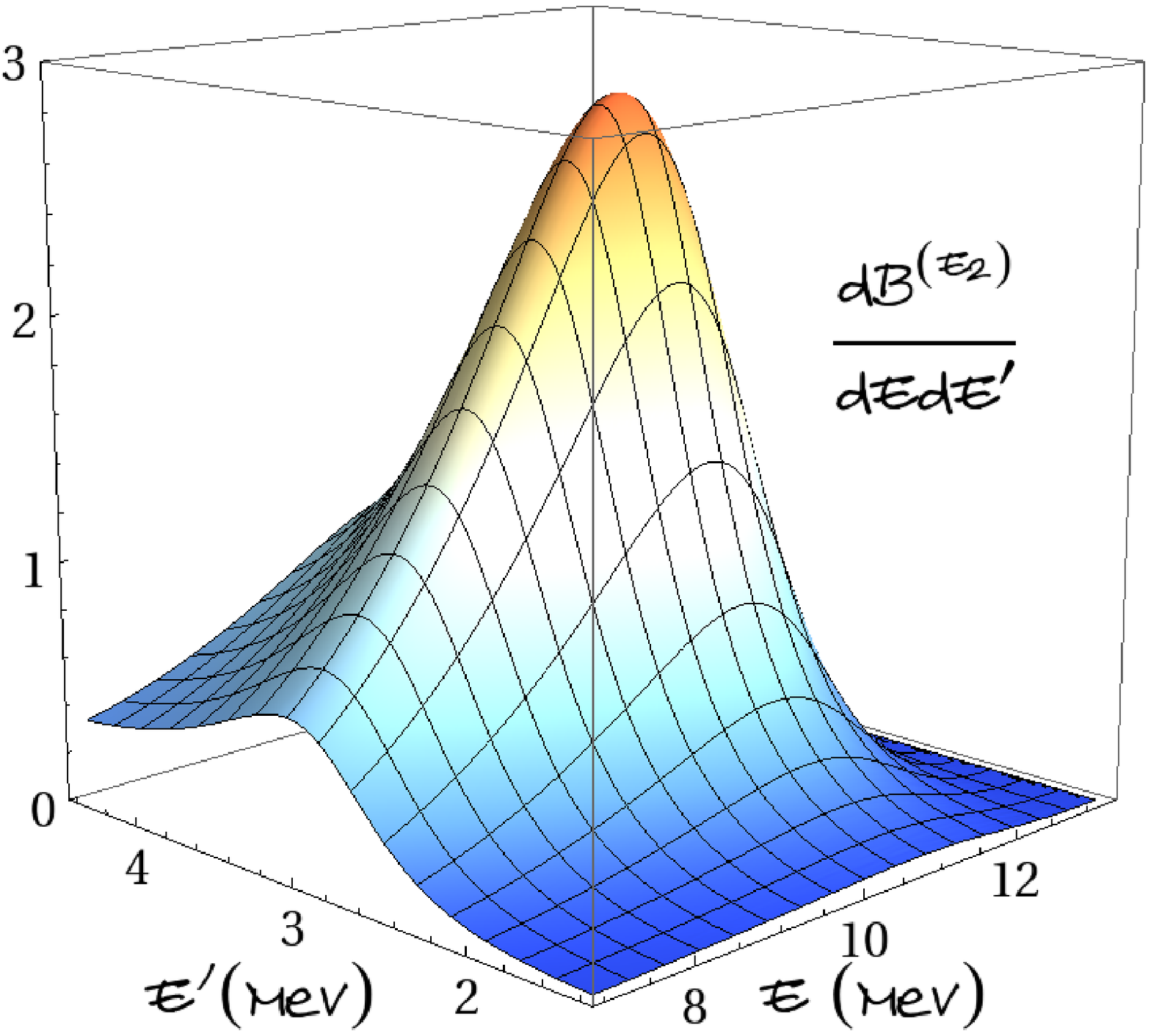,width=7.0cm,angle=0}
\caption{(Color online) Contour plots of the transition strength in
  Eq.(\ref{dbde}) as function of initial, $E$, and final state, $E'$,
  energies. The units are $e^2$~fm$^4/$MeV$^2$. The Buck potential is
  used. Upper and lower parts are for $2^+ \rightarrow 0^+$ and $4^+
  \rightarrow 2^+$, respectively.  }
\label{fig2a}
\end{figure}

The second method constructs $d{\cal B}^{(E2)}/dE dE'$ by dividing the
differential cross section in Eq.(\ref{eq7}) by the average value of
the photon energy and by the remaining constant factors. Integration
of the differential transition strength around the peak of the
resonance provides the total transition strength that will be denoted
as ${\cal B}^{(E2)}_\sigma$.  In Fig.\ref{fig2a} we show the $d{\cal
  B}^{(E2)}/dE dE'$ strength for the $2^+ \rightarrow 0^+$ and $4^+
\rightarrow 2^+$ transitions.  The distribution for the first case is
a very thin slice of the given final state energy along the initial
energy, both directions extending roughly as far as the respective
resonance widths.  This is for the same reason reflected in the
contour plot of the much broader $4^+ \rightarrow 2^+$ transition.

The transition strengths obtained with these two methods depend on the
energy window chosen around the resonance energy in the final
state. This window defines the integration range for $E'$ in
Eq.(\ref{eq4a}).  In Ref.\cite{gar13} the details about these two
methods are given, as well as the transition-strength values obtained
with them for different final energy windows. We have also found that
the computed strengths are insensitive to the two-body potential used,
and, for this reason, from now on only the results obtained with the
Buck potential will be given.

In table~\ref{tab2} we have collected the results obtained in \cite{gar13} for the 
$2^+\rightarrow 0^+$, $4^+\rightarrow 2^+$, $6^+\rightarrow 4^+$, and $8^+\rightarrow 6^+$
transitions for final energies $E'$ within the windows $E'_r \pm \Gamma'_r/2$ and
$E'_r \pm \Gamma'_r$, where $E'_r$ is the resonance energy in the final channel 
and $\Gamma'_r$ its corresponding width. These results are in fairly good agreement with
the ones obtained in \cite{lan86,lan86b} (8$^{th}$ column in the table).
However, the strength obtained for the $2^+ \rightarrow 0^+$ transition 
in Quantum Monte-Carlo calculations \cite{wir00} is clearly smaller, although
similar to \cite{gar13,lan86,lan86b} for the $4^+ \rightarrow 2^+$ case 
(9$^{th}$ column in the table).  The results shown for the $4^+ \rightarrow 2^+$,
transition are consistent with the experimental value of $25 \pm 8$ $e^2$fm$^4$ quoted in
\cite{dat05}.

\begin{table*}
\caption{${\cal B}^{(E2)}$-values (in $e^2$fm$^4$) for the different
  possible $E2$-transitions between the $^8$Be resonances. Columns from
  two to five are the results obtained in \cite{gar13} with the two methods 
  described in the text for a final energy window $E'_r\pm \Gamma'_r/2$ 
  (2$^{nd}$ and 3$^{rd}$ columns) and  $E'_r\pm \Gamma'_r$ (4$^{th}$ and 5$^{th}$ 
  columns), where $E'_r$ and $\Gamma'_r$ are the energy and width of the 
  resonance in the final state. The next two columns are the results obtained
  assuming a rotational model, Eq.(\ref{eq15}), when $Z_0$ is fixed to 3 fm 
  (6$^{th}$ column) and when $Z_0$ is taken equal to $\mbox{Re}\sqrt{\langle r^2 \rangle}$
  for each resonance (7$^{th}$ column). The results within parenthesis are the 
  ratios ${\cal B}^{(E2)}(\ell \rightarrow \ell')/{\cal B}^{(E2)}(2^+ \rightarrow 0^+)$
  for each of the calculations. The results from previous calculations 
  \cite{wir00,lan86,lan86b,har72} are given in columns 8 to 10. The last
  column is the result obtained after a complex scaling calculation assuming 
  a resonance to resonance transition.  }
\label{tab2}
\begin{ruledtabular}
\begin{tabular}{|c|cccc|cc|ccc|c|}
\hline 
${\cal B}^{(E2)}$ & \multicolumn{2}{c}{$E'_r\pm\Gamma'_r/2$} &
\multicolumn{2}{c|}{$E'_r\pm\Gamma'_r$} & \multicolumn{2}{c|}{Rotational model} & \cite{lan86,lan86b} &\cite{wir00} & \cite{har72} & Comp. scaling. \\  
   & ${\cal B}^{(E2)}_\gamma$ & ${\cal B}^{(E2)}_\sigma$ & 
     ${\cal B}^{(E2)}_\gamma$ & ${\cal B}^{(E2)}_\sigma$ & $Z_0=3$ fm & $Z_0$ &  &  &   &   \\ \hline
$2^+ \rightarrow 0^+$ & 53.4 (1)  & 32.9 (1)  & 79.1 (1) & 48.4 (1) & 6.4 (1) & 84.0 (1) & 71.3 & 14.8 & 16.8 & $-4.6 +i 33.6 (1)$ \\
$4^+ \rightarrow 2^+$ & 15.5 (0.29) & 12.1 (0.37)& 22.1 (0.28) & 17.2 (0.36) & 9.2 (1.43)&18.1 (0.22)  & 18.0 & 18.2 & 25.9 & $2.1+i11.8 (0.336-i0.0483)$ \\
$6^+ \rightarrow 4^+$ & 6.7 (0.13) & 4.5 (0.14) & 10.1 (0.13) & 6.9 (0.14) &10.1 (1.57) &9.1 (0.11) & - & -   & 33.9 & $3.0+i12.6(0.356-i0.139)$ \\ 
$8^+ \rightarrow 6^+$ & 6.6 (0.12) & 2.5 (0.08) & 13.0 (0.16) & 5.2 (0.11) &10.6 (1.65) &7.6 (0.09) & -&-  & - & $-6.1+i 13.2(0.410+i0.124)$ \\ 
\hline
\end{tabular}
\end{ruledtabular}
\end{table*}

An additional, and in a sense decisive, test of the rotational character
of the states in $^8$Be is provided by the total transition strength
given in Eq.(\ref{eq13a}).  For rotational bands with an inert
intrinsic structure, the total strength for two inert
$\alpha$-particles at $\pm Z_0/2$ is \cite{bohrII}
\begin{eqnarray} 
{\cal B}^{(E2)}(\ell \rightarrow \ell^{\prime}) = 
  \frac{5e^2}{4\pi} Z_0^4 \langle \ell 0 2 0  |\ell^{\prime} 0  \rangle^2 \;.
\label{eq15}
\end{eqnarray}
The spatial extension of the spherical $\alpha$-particle distribution
does not enter this expression in contrast to the moment of inertia in
Eq.(\ref{eq4}).  The different transition strengths are then related
by the expression:
\begin{equation}
\frac{ {\cal B}^{(E2)}(\ell_i \rightarrow \ell^{\prime}_f) } 
     { {\cal B}^{(E2)}(\tilde{\ell}_i \rightarrow \tilde{\ell^{\prime}}_f) }=
\frac{ \langle \ell_i 0 2  0  |\ell^{\prime}_f 0  \rangle^2}
     { \langle \tilde{\ell}_i 0 2 0  |\tilde{\ell^{\prime}}_f 0  \rangle^2} \;.
\label{eq5}
\end{equation}
The approximation in Eq.(\ref{eq5}) is valid for rigorous rotational
bands, and therefore in particular also for two rotating
$\alpha$-particles where Eq.(\ref{eq15}) applies.  Comparing to
Eq.(\ref{eq13a}) this is seen to imply that the integrals should be
independent of the transition, which reflects that the radial wave
functions and then the intrinsic structure is the same for all the
states.  

In the schematic rotational model of Eq.(\ref{eq15}) we get
all the transition strengths for a given $Z_0$.  They are useful for
comparison and interpretation. We first choose a constant $Z_0 = 3$~fm, 
which was the value chosen to obtain the sequence of states denoted by 
$E_r^{(1)}$ in table~\ref{tab1}. The strength values given by
Eq.(\ref{eq15}) are shown in the 6$^{th}$ column of table~\ref{tab2}.
They are clearly different to the ${\cal B}^{(E2)}$-values obtained in
\cite{gar13}, no matter the size of the window used and the procedure 
used to extract it. It is quite clear that the transition strengths do not
follow the rule dictated by the strict rotational model.  

The same conclusion is reached when examining the transition strength
ratios.  The value of $\langle \ell_i 0 2 0 |\ell^{\prime}_f 0
\rangle^2$ is 0.2, 0.29, 0.31, and 0.33 for the $2^+ \rightarrow 0^+$,
$4^+ \rightarrow 2^+$, $6^+ \rightarrow 4^+$, and $8^+ \rightarrow
6^+$ transitions, respectively.  When taking the $2^+\rightarrow 0^+$
transition as a reference, the ratios given by the rotational model,
Eq.(\ref{eq5}), are shown by the numbers within parenthesis in the
6$^{th}$ column of the table.  The last three transitions should then
have a rather similar strength, which in turn should be larger than
the strength corresponding to the $2^+ \rightarrow 0^+$.  Nothing of
this happens with the $\alpha-\alpha$ potentials.  The ratios obtained
with the transition strengths in Ref.\cite{gar13} (given by the
corresponding numbers within parenthesis in each of the columns in
table~\ref{tab2}) are clearly smaller, and the maximum transition
strength is actually obtained for the $2^+ \rightarrow 0^+$
transition.  

The behavior predicted by the rotational model coincides with the one
found with the microscopic cluster model \cite{har72} (10$^{th}$
column in the table), although the absolute values are about a factor
of 3 different. This corresponds to a larger value of $Z_0 \approx
4$~fm consistent with the spatial extension found in \cite{har72}.
This resemblance of the rotational model and the classical microscopic
cluster model results is perhaps not very surprising, since the
$\alpha-\alpha$ structure after all is imposed in both cases.

However, the cluster model in \cite{har72} is based on a generator
coordinate description where angular momentum projection before and
after variation both start out with the same $\alpha-\alpha$ cluster
structure.  The different angular momentum states are then related
through a similar intrinsic structure, which can be somewhat
differently deformed depending on angular momentum, but still the
basic rotational model assumptions are approached and almost
fulfilled.  In contrast, the potential models with effective
$\alpha-\alpha$-interactions provide independent solutions for each of
the angular momenta.  The solutions are only related through the
same central potential.

It is then clear that the radial integrals in Eq.(\ref{eq13a}) change
with angular momentum and produce unexpected transition strengths.
The different spatial structure of the resonances was already seen
when analyzing the $\langle r^2 \rangle^{1/2}$-values, which for
instance for the $0^+$ case is about twice the value in the $4^+$,
$6^+$, or $8^+$ cases.  In fact, in the previous section we saw that
when using different values for $Z_0 =\mbox{Re}\sqrt{\langle r^2
  \rangle}$ for each resonance, the energy sequence in $^8$Be was
nicely reproduced. It is then very tempting to check if the same good
agreement is recovered when using $Z_0 =\mbox{Re}\sqrt{\langle r^2
  \rangle}$ for each resonance in Eq.(\ref{eq15}). More precisely we
have chosen for each transition the $Z_0$ value corresponding to the
final state resonance. The results obtained are given in the 7$^{th}$
column of table~\ref{tab2}. As we can see, now the agreement with the
results in \cite{gar13} is definitely much better, especially with the
${\cal B}^{(E2)}_\gamma$-values when using the $E'_r\pm \Gamma'_r$
final energy window. As a consequence, the corresponding ratios
(numbers within parenthesis) also agree much better now. This
result seems to confirm the conclusion reached in the previous
section, namely, the $^8$Be spectrum has a rotational character
provided that the $\alpha-\alpha$ distance is angular momentum
dependent. Still the principal $\alpha$-cluster structure is
maintained.

\section{Discussion}

To discuss quantitatively we should preferentially apply the method to
specific systems as we did in the previous sections.  We shall here
first discuss the radial dependence of wave functions in the
continuum.  This is numerically simple by use of the complex scaling
method.  However, the properties of the corresponding complex rotated
wave function then only represent a part of the cross section.  Other
parts related to continuum contributions are necessary to obtain the
full observable cross sections. Complex scaling mixes these
contributions in a complicated manner, but we expect the resonance
structures to be strongly indicative for the overall behavior.

To supplement we discuss instead the properties of the wave functions
and the resulting transitions for real energies, where the interesting
physics is hiding behind diverging integrals.  We continue to discuss
basic conditions for appearance of rotational motion in two-body
systems.  We then turn high-spin states created in heavy-ion
collisions often claimed to be of rotational structure.

\subsection{Continuum resonance structures}

The definition of transition probabilities is in practice not well
defined since the states are not well defined either in the continuum.
It is then interesting to know the results for transitions between the
rigorously defined resonance states found by complex rotation.
However, the probabilities are then ``rotated'' into the complex
values given in the last row of table~\ref{tab2}.  The results in the
present work do not depend significantly on the potential, and we
therefore only give the results for the Buck potential.  They are
independent of rotation angle as required by well defined resonances,
but obviously they cannot represent observable quantities.  First, the
results are complex numbers. Second, they are only part of the full
observables, which include both resonance-to-resonance and continuum
background contributions \cite{gar13a,myo98}.

The ratios of these partial transition probabilities (in brackets in
the last column in table~\ref{tab2}) now show the same behavior of the
factor of $3$ decrease from the $2^+ \rightarrow 0^+$ to the $4^+
\rightarrow 2^+$ transition as for the full calculation using only
real energies. The imaginary part is 10 times smaller than the real
part.  However, the real parts of the next two ratios involving the
rather artificial $6^+$ and $8^+$ resonances increases almost in line
with the schematic rotational model.  The imaginary parts of these
complex numbers are still a factor of 3 smaller than the real parts of
these ratios.

Therefore, the observable relative transition probabilities
for the first 3 resonances (not the last) in the first columns of
table~\ref{tab2} are almost recovered in the complex
resonance-to-resonance relative transitions.  These can in turn be
understood by their decreasing $\sqrt{\langle r^2 \rangle}$-values of
the resonances as given in table~\ref{tab1}.  Thus, we can conclude
that the very large deviations from the rotational model arise from a
decreasing spatial extension of the $0^+$, $2^+$ and $4^+$ resonances.

\begin{figure}
\epsfig{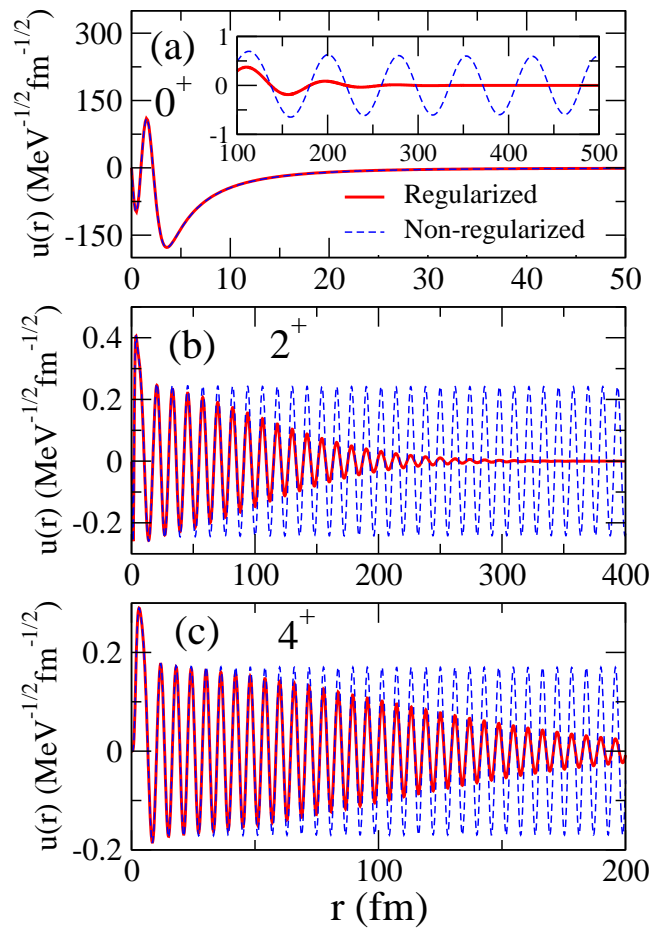}
\caption{(Color online) The radial wave functions for the three lowest
  resonances, $0^+$, $2^+$, and $4^+$, are shown as function of $r$.
  The solid and dashed curves correspond to the regularized and
  non-regularized wave functions, respectively. The regularization has
  been performed by introducing the Zel'dovich factor
  $\exp(-\eta^2r^2/2)$, with $\eta =0.01$~fm$^{-1}$.  The energy of each
  wave function is taken at the center of the resonance, $E_r$. }
\label{fig3a}
\end{figure}

However, an understanding of the properties obtained entirely by
calculations for real energies is much more complicated.
Nevertheless, we shall in the following attempt a detailed
explanation.  The radial wave functions for the lowest three
resonances are shown in Fig.\ref{fig3a} both non-regularized and
multiplied by an appropriate Zel'dovich factor. 
In principle, for a given value of the Zel'dovich
parameter, the radial wave functions should be accordingly renormalized, such 
that Eqs.(\ref{ener}) or (\ref{ort}) are restored. However, 
the wave functions are already initially normalized, and
when taking the limit of zero $\eta$-value the correct normalization condition
is recovered faster than the converged value of the radial integral. Therefore,
in practice the renormalization is not needed.

As seen in the figure, the $0^{+}$ state
behaves as a bound state, although tiny oscillations are visible at
large distances before the Zel'dovich cut-off becomes efficient (inset in 
Fig.\ref{fig3a}a).  The
two nodes at small distances are due to the deep potential with two
spurious strongly bound states.

For both the $2^{+}$ and $4^{+}$ resonances the oscillations are very pronounced up
to about $200$~fm, where the Zel'dovich regularized wave function
essentially has vanished.  The resonance structures are only visible
at very small distances, where the first oscillation of each state
inside the attractive region of the potential, has twice the amplitude
of the second. Note that in Fig.4 only the
radial wave functions $u(r)$ are shown, while the total radial
wave function is actually given by $u(r)/r$. When dividing by $r$ we can
see that the $0^+$ resonance behaves as a bound state, and both, the $2^+$ and
$4^+$ states, reveal resonance character only at distances smaller than about $5$~fm.
The period in the oscillations in Fig.\ref{fig3a}
depends only on the resonance energies through the wave number,
$k=\sqrt{2\mu E/\hbar^2}$, which gives wave lengths $2\pi/k \approx
70$~fm, 12~fm, and 6~fm, respectively.  The Coulomb barriers for these
states extend correspondingly to about 60~fm, 10~fm, and 5~fm, that is
the regular oscillations all occur outside the barriers for positive
kinetic energies.

\begin{figure}
\epsfig{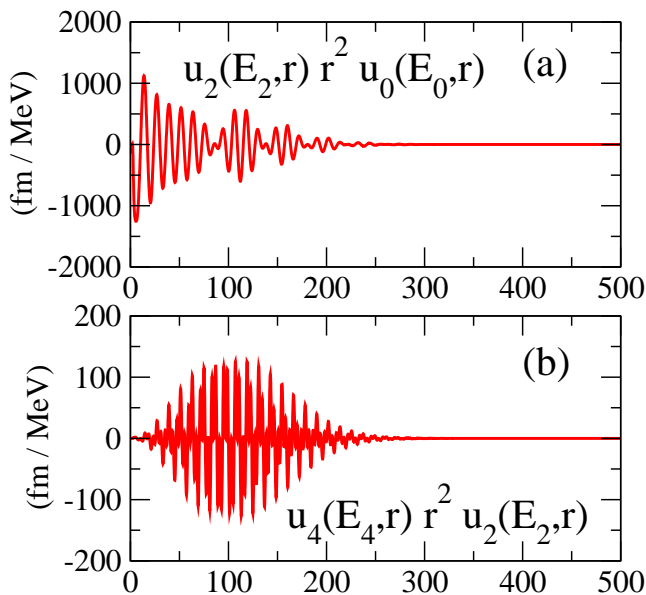}
\caption{(Color online) The radial integrands of the two lowest transitions, $2^{+}
  \rightarrow 0^{+}$ and $4^{+} \rightarrow 2^{+}$ as function of $r$
  after Zel'dovich regularization with the factor (squared) in
  Fig.\ref{fig3a}, and for the same energies.  }
\label{fig3b}
\end{figure}

The transitions are determined as integrals over radial wave
functions, see Eq.(\ref{eq13a}).  We show in Fig.\ref{fig3b} the
integrands of the matrix elements for the lowest transitions for
energies corresponding to resonance peaks.  The oscillations appearing
now are the results of combining the two oscillating wave functions.
Their different periods produce the different (from the wave
functions) but regular oscillation extending to about 200~fm.  For
$2^{+} \rightarrow 0^{+}$, the revival after destructive interference
is seen before the Zel'dovich cut-off reduce the amplitude to be
insignificant.  For $4^{+} \rightarrow 2^{+}$, the amplitude increases
to about 100~fm/MeV up to 100~fm and is in fact only a few fm/MeV at
small distances. 

\begin{figure}
\epsfig{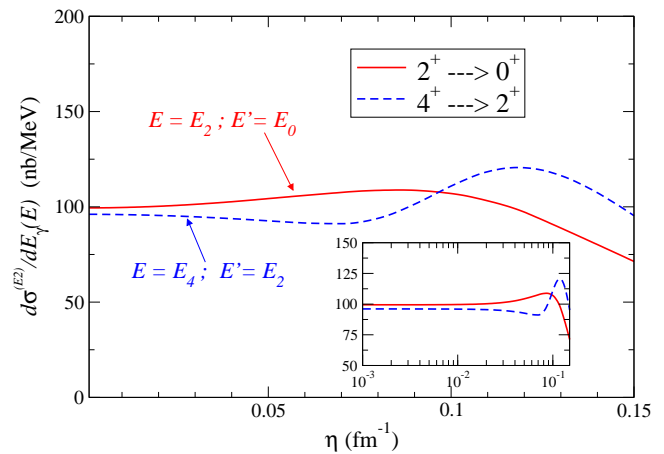}
\caption{(Color online) Differential cross section
  (Eq.(\ref{eq3})) for given initial and final energies ($E$ and
  $E^\prime$) for the $2^+ \rightarrow 0^+$ (solid curve) and $4^+
  \rightarrow 2^+$ (dashed curve) transitions in $^8$Be as a function
  of the Zel'dovich parameter $\eta$. The initial and final energies
  for each transition correspond to the $0^+$, $2^+$, and $4^+$ 
  resonance energies (same energies as in Figs.\ref{fig3a} and \ref{fig3b}).
  The inset shows the same as in the main figure but with the $\eta$-axis
  in logarithmic scale.}
\label{zeld}
\end{figure}

It is now highly significant that the integrals themselves are only a
few fm$^2$/MeV.  This means that the oscillations of the integrands
cancel to a very large extent, where it may be necessary to emphasize
that a substantial range of Zel'dovich parameters, $ \eta \lesssim
0.1$ fm$^{-1}$, produces precisely the same matrix element.
This is illustrated in Fig.6, where the solid and dashed curves show the 
integrals of the functions in Figs.\ref{fig3b}a and \ref{fig3b}b, but
as a function of the Zel'dovich parameter $\eta$. As we can see, for sufficiently small
values of $\eta$, the computed integrals become constant.
The wave functions and the matrix elements in
Figs.\ref{fig3a} and \ref{fig3b} are shown for $\eta = 0.01$~fm$^{-1}$,
which reveal the large amplitude oscillations at rather large
distances. A variation of $\eta$ from very small to large values,
$\eta \sim 0.01$~fm$^{-1}$, would move the damping of the wave
functions in Fig.\ref{fig3a} and the oscillating structures in
Fig.\ref{fig3b} down to smaller distances, bit still leaving 
untouched the small distance part of the wave functions, where
the resonance structure is contained. 

It is then remarkable that the transition probabilities between states
of given energies are numerically well defined to values much smaller
than corresponding to the large amplitudes at large distances.
However, to extract the decisive short-distance properties of the
resonance wave functions is much more difficult.  The cancellation at
large distances implies that these oscillations only play a minor role
in the determination of the transition probability.  In fact, only
distances of less than about $5$~fm contribute corresponding to
the spatial extension of the regions where resonance character is seen
in Fig.\ref{fig3a}.  To be on the safe side where the matrix elements
still can be reliably obtained numerically, we choose the value of
$\eta = 0.01$~fm${-1}$ in all cases investigated in the present work. 
More detail will be presented in ref.\cite{gar13a}.

If we use the complex scaling method, the oscillations in
Figs.\ref{fig3a} and \ref{fig3b} disappear altogether in the complex
scaled resonance wave functions, and radii and transition matrix
elements are well defined.  This does not prevent larger distances
from giving significant contributions.  The smaller radii for larger
angular momenta seen in table~\ref{tab1} are the opposite of the
ordinary centrifugal stretching.  This is related to properties of
real energy calculations where the increasing widths of the resonances
arise as they approach the top of the barriers.  The states then
approach free waves in most space. 

The free wave oscillations are quickly approached in Figs.\ref{fig3a}
and \ref{fig3b} before the Zel'dovich factor is applied.  Only
deviations from the free wave can contribute to resonance properties,
and in turn to results for transition probabilities like ${\cal
  B}^{(E2)}$-values.  Therefore, the smaller the radii of the space
exhibiting deviations from free waves, the smaller are the radial
moments, and in turn the radial transition matrix elements also
decrease.

\begin{figure}
\epsfig{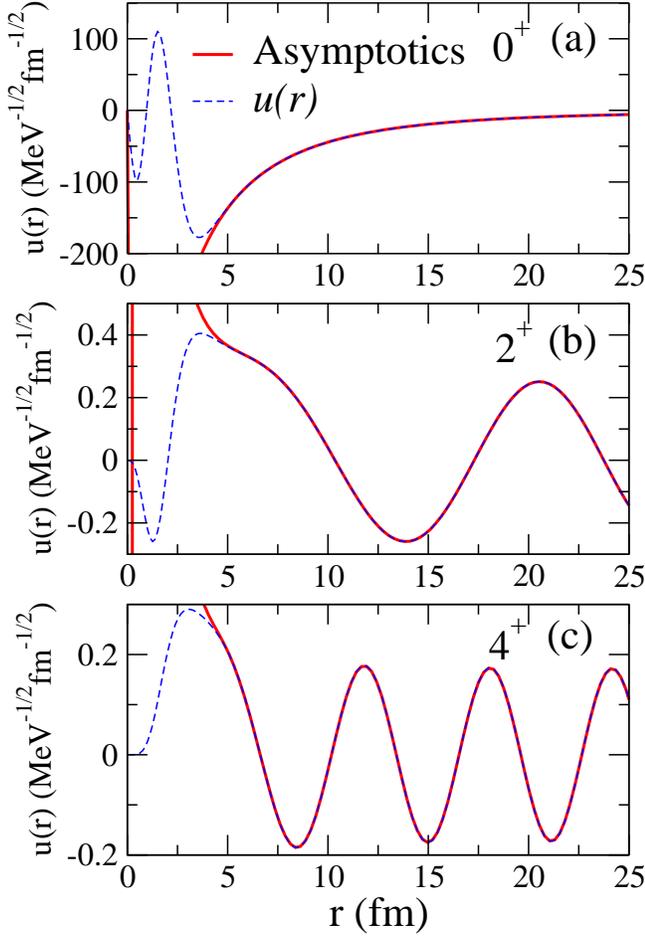}
\caption{(Color online) The solid curves show the asymptotic wave
  function, Eq.(\ref{asymp}), for the three lowest resonances, 
  $0^+$, $2^+$, and $4^+$. The corresponding radial functions 
  $u(r)$, already shown in Fig.\ref{fig3a}, are given by the dashed
  curves. The energy of each wave function is taken at the center of 
  the resonance, $E_r$.}
\label{fignew}
\end{figure}

To understand this a little better we turn to the asymptotic behavior
of the continuum wave functions in Eq.(\ref{asymp}).  Let us first
define the asymptotically vanishing function
\begin{eqnarray}
 \tilde{u}_\ell(E,r) =  u_\ell(E,r) - u_\ell^{asymp}(E,r) \; \label{smooth} \\
u_\ell^{asymp}(E,r) = \lim_{r \rightarrow \infty} u_\ell(E,r) \;. 
\label{smooth1}
\end{eqnarray}

The remaining function, $\tilde{u}_{\ell}$, now contains the resonance
structure revealed at short distances, and traces of the oscillating
continuum structures are removed. 
The precise asymptotic behavior from Eq.(\ref{asymp}) is  only correct for point-like charge 
distributions, or at distances where the charges do not overlap.  At smaller distances the function 
in Eq.(\ref{asymp}) is incorrect and even diverging for distances approaching zero. To get 
physically meaningful results we have to account for the finite extension of the charges. This is 
easily done by a regularization procedure or by extending the Coulomb wave functions down to 
zero by combinations of sine and cosine functions as in the case of no Coulomb interaction or 
from the corresponding asymptotic limit of the Coulomb functions. We choose first the true asymptotics 
from Eq.(\ref{asymp}) and show in Fig.\ref{fignew} the three resonance wave functions, $u$ (dashed
curves), and its asymptotic behavior (solid curves), as function of $r$. At distances larger than 
$5$~fm the full and the asymptotic wave functions are indistinguishable.
This is in full agreement with the resonance radii in Table~\ref{tab1}, Fig.\ref{fig3a}, as well as 
the discussion in connection with the transition matrix elements in Fig.\ref{fig3b}. 

The division into short and asymptotic parts in Eqs.(\ref{smooth}) and (\ref{smooth1}) 
is now directly applicable in a separation of contributions from the
different parts. This is highly desirable in analyses of experimental
data using the $R$-matrix formulation where any continuum contribution
appears as spurious resonances strongly depending on the channel
radius. 
We can calculate the radial transition matrix element, 
$B_{\ell,\ell'}(E,E')=\langle u_\ell(E,r)|r^2|u_{\ell'}(E',r) \rangle$, 
which naturally is divided
into four types of terms involving short-distance and asymptotic
parts in different combinations, i.e.
\begin{eqnarray}
\label{smootha}  && B^{(sh,sh)}_{\ell,\ell'} = 
\langle \tilde{u}_\ell(E,r)|r^2|\tilde{u}_{\ell'}(E',r) \rangle \;, \\
\label{smoothb} &&  B^{(sh,as)}_{\ell,\ell'}   = 
\langle \tilde{u}_\ell(E,r)|r^2|{u}^{asymp}_{\ell'}(E',r) \rangle  \;, \\
\label{smoothc} && B^{(as,sh)}_{\ell,\ell'}   = 
\langle {u}^{asymp}_\ell(E,r)|r^2| \tilde{u}_{\ell'}(E',r) \rangle  \;, \\
\label{smoothd} &&  B^{(as,as)}_{\ell,\ell'}  = 
\langle {u}^{asymp}_\ell(E,r)|r^2| {u}^{asymp}_{\ell'}(E',r) \rangle  \;, \\
\label{smoothe} && B_{\ell,\ell'}  =  B^{(sh,sh)}_{\ell,\ell'} +  B^{(sh,as)}_{\ell,\ell'}
                                        + B^{(as,sh)}_{\ell,\ell'} + B^{(as,as)}_{\ell,\ell'}\; ,
\end{eqnarray}
where the notation of the contributions to $B_{\ell,\ell'}$ refers to
short-distance and asymptotic combinations.  A tempting interpretation
is the correspondence of contributions from respectively resonance to
resonance $(sh,sh)$, resonance to continuum $(sh,as)$, continuum to
resonance $(as,sh)$, and continuum to continuum $(as,as)$.

At very large distances the wave functions $u_\ell$ are governed by a 
combination of sine and cosine functions of $(\kappa r)$, as it can be
seen from Eq.(\ref{asymp}). The terms of the type $B^{(as,as)}_{\ell,\ell'}$
contain then in the integrand products of either two sine functions, or two
cosine functions, or one sine and one cosine functions. These integrals are
therefore not well defined unless the Zel'dovich regularization is applied.
When done, we obtain vanishing
results both when two $\sin(\kappa r)$-type and two $\cos(\kappa r)$-type 
functions are combined, whereas finite results emerge when products of $\sin(\kappa r)$ 
and $\cos(\kappa' r)$ functions appear (or the other way around). These conclusions 
assume that $E \neq E'$.
All other terms in $B_{\ell,\ell'}$ have the $\tilde{u}$-functions as factors, and the
corresponding integrals are convergent and well defined.

\begin{figure}
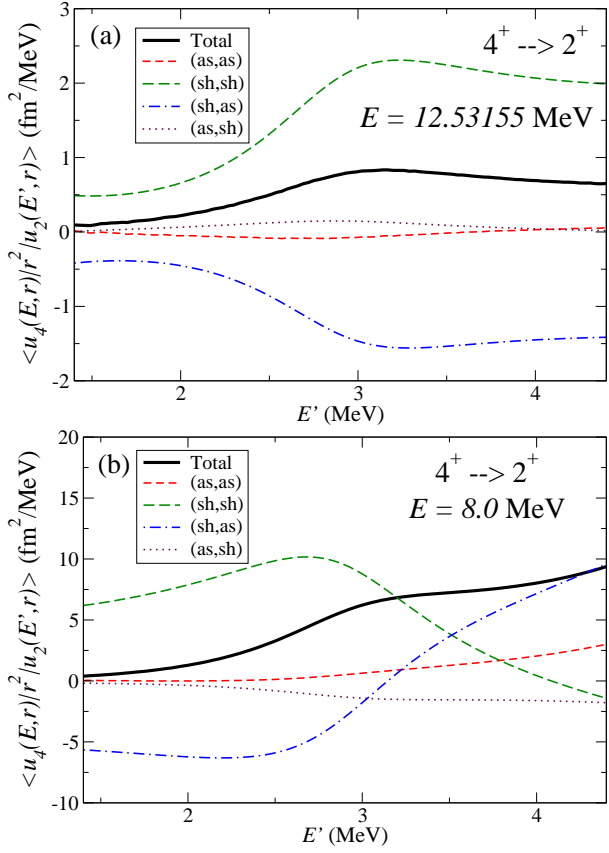

\epsfig{file=parts-3.eps,width=8.0cm,angle=0}
\epsfig{file=parts-2.eps,width=8.0cm,angle=0}
\caption{(Color online) Radial transition matrix element
  $B_{\ell,\ell'}(E,E')$ for the $4^+\rightarrow 2^+$ transition as a
  function of the final energy $E'$.  The total transition matrix
  element is given by the solid curve. The $(as,as)$, $(sh,sh)$,
  $(sh,as)$, and $(as,sh)$ contributions are shown by the
  short-dashed, long-dashed, dot-dashed, and dotted curves,
  respectively.  In part (a) the initial energy is $E=12.53155$ MeV
  for which the $4^+$ phase-shift is equal to $\pi/2$. In part (b) the
  initial energy ($E=8$ MeV) has been chosen to be outside the $4^+$
  resonance peak.  }
\label{figp}
\end{figure}

One immediate consequence is that $B^{(as,as)}_{\ell,\ell'}$ is zero
when initial and final state wave functions both correspond precisely
with resonance states, where the phase shifts are $\pi/2$. The
products of the asymptotic parts are equal to zero since
$\cos(\pi/2)=0$, and only terms of the type $\cos(\kappa r)$ survive
in the asymptotic part of the wave function.  Therefore, only terms
where the $\tilde{u}$ wave functions enter give non-vanishing
contributions to $B_{\ell \ell'}$. This is illustrated in
Fig.\ref{figp}a, where we plot $B_{4^+,2^+}(E,E')$ as a function of
the final energy $E'$ and for an initial energy of $E=12.53155$ MeV,
which corresponds to a $4^+$ phase-shift equal to $\pi/2$. The total
value is shown by the solid curve. The
$B^{(as,as)}_{4^+,2^+}$-contribution is given by the short-dashed
curve, which is very small for all the final energies shown, and it is
particularly close to zero in the vicinity of $E'=3.33$ MeV, which is
the value at which the 2$^+$ phase shift is $\pi/2$.

This is consistent with our classical intuition where resonances are
located in the continuum with large amplitudes at short distances and
comparably small and non-contributing amplitudes at large distance.
Still, even here the terms $B^{(as,sh)}_{\ell,\ell'}$ and
$B^{(sh,as)}_{\ell,\ell'}$ are non-vanishing, although only relatively
short distances can contribute due to the $\tilde{u}$-functions.
These terms may represent unavoidable continuum background
contributions as it can be seen in Fig.\ref{figp}a, where especially
the $(sh,as)$ contribution (dot-dashed curve) is very significant.
The interference between the different contributions is very
substantial.  However, the contributions must arise from radii where
the $\tilde{u}$ functions are finite, that is the large amplitude
oscillation seen in Fig.\ref{fig3b} necessarily cancels completely.

When the initial and final state energies differ from the resonance
energies the $B^{(as,as)}_{\ell,\ell'}$-contribution is then not
identically equal to zero, although well defined after the Zel'dovich
regularization. This is shown in Fig.\ref{figp}b, where the initial
energy ($E=8$ MeV) does not correspond to any $4^+$ resonance.  The
interference contribution is again substantial.  In this case the $(as,as)$
contribution (short-dashed curve) is clearly relevant, although the
major contribution from the asymptotic part is again the $(sh,as)$
one, given by the dot-dashed curve.

We are now equipped to summarize the validity of the ordinarily used
long wavelength approximation where the contributing radii should be
smaller than the inverse of the wave number.  The total resonance wave
function in Fig.\ref{fig3a} reveals a larger amplitude at small distances than
for the asymptotic oscillations.  This is much clearer seen in Fig.\ref{fignew},
where the deviation between full and asymptotic wave functions are
shown.  The radial extension is in agreement with the root-mean-square
values of the resonances given in Table~\ref{tab1}.  The contribution to the
matrix elements arise from rather small distances as seen by comparing
amplitudes of the integrand in Fig.\ref{fig3b} with the very much smaller
integrated result.  The large large-distance oscillations must
therefore essentially cancel. 

The non-vanishing values of $\tilde{u}$ necessary to get
contributions are confined to radii less than  5 fm as seen in
Fig.\ref{fignew}.  This is much smaller than the smallest contributing
wavelength of $25$~fm arising from the largest contributing photon energy of
$10-12$~MeV. The estimate of the contributing photon energy interval
can be seen in Fig.\ref{fig2a}, where we show the distribution of strength for
the transition cross sections from given initial to final state
energies.  The bulk contribution are concentrated in peaks
corresponding to a photon energy of substantially less than $12$~MeV in the worst
case of the tail of the $4^+ \rightarrow 2^+$ transition. Thus, the long wavelength
approximation is rather accurate.

\subsection{Validity of the rotational model}

The classical rotation of an isolated inert system is characterized by
its kinetic energy, which can be expressed as either the square of the
angular momentum divided by twice the moment of inertia around the
rotation axis, or half of the square of the rotation frequency times
the same moment of inertia.  

The convenient quantum mechanical version is in terms of the conserved
and quantized angular momentum.  An ideal analogue is then a two-body
cluster structure with a strongly attractive one-dimensional
delta-shell potential like $\delta(r-Z_0)$.  The ground state wave
function of zero angular momentum is localized at the relative
distance $r=Z_0$. This corresponds to an intrinsic wave function
localized in one point, $z=Z_0$, and averaged with equal weights over
all spatial directions.  For finite angular momentum, $\ell$, the
energy is increased by $\hbar^2\ell(\ell+1)/(2{\cal I})$, where ${\cal
  I} = \mu Z_0^2$.  The relative distance for a bound state would
still be $Z_0$, and the use of relative coordinate requires the use of the
reduced mass $\mu$.  The rotational spectrum is then recovered.

For an attractive potential of finite range like a square well or a
gaussian potential, the rotational structures do not automatically
appear.  A repulsion at short distance and a confining barrier at
larger distances would lead to a potential of $\delta$-shell
character.  If it is deep enough the rotational spectrum would then
again arise.  These repulsive potentials could in principle be
provided by a Coulomb interaction between point-like particles. Both
short and long-range repulsions would appear.  However, these barriers
are very easily either small or not present at all.

Conditions for a rotational spectrum with only an attractive finite
range potential can be seen by use of simple potentials.  The harmonic
oscillator potential gives energies linear in $\ell$ while the
energies for a spherical square well potential are more promising.  For
a deep-lying bound state the boundary condition is approximately,
$j_{\ell}(\kappa R) = 0$, where $j_{\ell}$ is the spherical Bessel
function of order $\ell$, $R$ is the radius and $\kappa$ the wave
number.  The nodes of these Bessel functions approach
$(\ell+1/2)(1+1.86/(\ell+1/2)^{2/3})$ for large $\ell$, and the
corresponding energies found from the square of $\kappa$ are then
approaching $(\ell+1/2)^2$ which is the semiclassical analog of
$\ell(\ell+1)$.  Thus in this limit the rotational spectrum also
emerges, and three levels of the rotational sequence can approximately
be reproduced with one free parameter like the radius or the moment of
inertia. Then, we conclude that  if a flat potential is sufficiently deep then the 
rotational energy sequence arises. This reflects the classical knowledge 
that the kinetic energy is responsible for the rotational character of 
a system described by a rotational invariant hamiltonian. The other limit, 
where the potential is unable to support bound states of non-zero or 
moderate angular momenta, is for the same reason not necessarily of 
rotational character. It is then somewhat surprising that the $^8$Be states 
to some extent reveal this character even as resonances.

For states deeply bound in a short-range attractive potential the
centrifugal barrier term is comparatively small and the radial wave
functions are expected to be roughly independent of $\ell$.  Then the
rotational sequence of transition probabilities would be approximately
obeyed. However, these states must be strongly bound, and certainly
not simply unbound resonance structures in the continuum.  We can then
conjecture that the rotational model for a two-body inert structure is
valid for very strong short-range attractions, and viceversa invalid
when the attraction becomes comparable to the centrifugal barrier term.

It is still not excluded that resonance structures approximately could
obey the necessary independence of the radial integrals in order to
validate the rotational transition sequences. The regularization of
the continuum wave functions influences the corresponding radial
integrals but it is not obvious that the higher-lying energy and
angular momentum states are less spatially extended.  The explanation
is that the oscillating behavior appears at shorter distances when the
state is closest to the barrier.  The regularization removes the
corresponding contribution by subtraction of the diverging
large-distance part of the wave function. The result is decreasing
radial integral and increasing deviation from the rotational model.
This could be an artifact of the present procedure but the measured
datum confirms this interpretation.  It is therefore highly
interesting to obtain more experimental data for verification or
possibly falsification of the present interpretation.

We should finally emphasize that the presence of other types of
excitation also may destroy the validity of the pure rotational model.
Such bound states or resonances are abundantly arising from intrinsic
degrees of freedom or other collective motion like vibrations.
Effective decoupling of rotational states from other excited states
would be achieved when the excitation energies of the rotations are
much lower than all other excitations.  However, this can not continue
indefinitely to high excitations where other degrees of freedom may be
excited as well.  For comparable energies the coupling producing mixed
states can be very moderate and the pure pictures are no longer valid.
It is still possible to have rotational states at relatively high
energy provided that other degrees of freedom either produce excited states
far away and of much larger energy separation than the
rotations, or in practice decoupled due to e.g. disparate structures of all 
other excited states of comparable excitation energies.

\subsection{Rotational states of heavy nuclei}

The fact that a large number of nuclear spectra exhibit rotational
structure \cite{bohrII,cas00} demands an explanation.  The heavy-ion
populated high-spin states reach more than 50 units of $\hbar$.  Many
different high-spin bands apparently appear in the same nucleus, a
typical example can be found in \cite{ril88}. Transitions are measured
between intra, as well as inter, band members.  Still, the
interpretation is in terms of rotational bands.

As discussed above, the validity of the rotational model seems to rely
on an effectively strong binding which allows the exited states to be
strongly bound as well.  This is achieved for nuclear states where
lifetimes or widths are determined or dominated by photon emission
processes.  Any other decays like fission, nucleon and cluster
emission are then strongly hindered.  A barrier must then effectively
be present in all other decay channels than photon emission.  The
result is that the nucleus then must behave as a strongly bound
system.

This apparent strong binding can be directly due to a huge barrier
against decay as for example fission of intermediate mass nuclei. A
barrier may also be effectively present if restructuring is required
to arrive at the final decay product as for example for
$\alpha$-emission of nuclei without traces of $\alpha$-clustering.

The pronounced rotational structures are also first of all found for
relatively small energies.  The numerous high-spin states and the
abundantly experimentally obtained rotational spectra are not
necessarily contradicting this interpretation. 

The transition probabilities for the high-spin states are also often
not following the rotational model that well. There is always the
centrifugal stretching, higher order corrections even to the
rotational energy spectra, deformation and pairing variations,
etc. \cite{bohrII,cas00}. Furthermore, if the preferred decay channel
is fission, particle or cluster emission, the large width of the
states prohibit accurate direct measurement of the transition
probabilities. Such states may possibly be members of a rotational
sequence of energies, but their photon emission probabilities are not
observables.  A full population and decay history in terms of cross
sections are required to get a meaningful description as in the case
of $^8$Be discussed in this paper.

\section{Summary and conclusion}

We investigate the simplest structure able to exhibit quantum
mechanical rotational motion: two spin-zero inert $\alpha$-particles.
We first sketch the formalism which is precisely valid for bound
states, but present increasing problems as the continuum properties
becomes more pronounced.  We first explain that it is absolutely
necessary to use observables for continuum calculations.  The
structure information via the ${\cal B}^{(E2)}$-values cannot be
obtained directly without very severe restrictions to energies around
the resonances between which the transition occurs.

Staying with observables has the positive implication that direct
comparison with measurements is possible.  However, the structure
information is then hidden in pieces of the observables.  One
conclusion is therefore that structure and reaction cannot be
disentangled and we have to live with this lack of information about
the continuum structures. We show that it is possible to derive
structure information, although the results are inherently uncertain
due to the unavoidable use of non-observable quantities.  We describe
two procedures to derive the non-observable ${\cal B}^{(E2)}$-values
which contain information about the structure of the resonances.

We show that the rotational energy sequence and corresponding radii
are followed by the resonances. However, in contrast the transition
probabilities deviate substantially from the rotational model
predictions. First, this is not due to the uncertainties arising from
the extraction of these non-observable continuum properties. It is
also not due to neglect of intrinsic $\alpha$-structure,
$\alpha$-polarization or effective charges, or centrifugal stretching
effects.  The deviations are traced to an unexpected radial dependence
of the relative resonance wave functions.  They contract as the
barrier is approached, and the only experimental point confirms this
result.  However, it is worth emphasizing that the corresponding
continuum wave functions are a priory non-normalizable. A suitable
regularization procedure is necessary to extract observable quantities,
which in turn can be related to the mentioned radial contraction.

In classical cluster models the rotational predictions are followed
much more accurately, as these models resemble rigid rotors.  Modern
variational or shell model calculations are often treating the
resonances as bound states. These calculations therefore altogether
unphysically avoid the problems connected to continuum properties.
The results are an uncontrolled average comparable to those of proper
continuum models but the tendencies do not point in one direction.

Our results are independent of the potentials employed as long as the
low-energy $\alpha-\alpha$ phase shifts reproduce the measured values.
This is somewhat surprising as the transition operator seems to be
sensitive to the contributing short-distance properties of the wave
functions.  The potentials are only marginally able to hold
resonances, and for example $\ell=4$ is even higher than the barrier
but still clearly revealing a pole in the $S$-matrix. This is,
strangely enough, also the case for $\ell=6,8$ where the widths are
huge and normally would be contradicting a description as resonance
states. A better interpretation is in terms of a broad background
contribution at these energies.

The many known low-energy rotational states in intermediate and heavy
nuclei presumably require no new interpretation.  They effectively
behave as bound states since they are below separation thresholds or
an enormous restructuring is required to decay through other channels
than photon emission.  We expect this cannot also hold for the
high-lying high-spin states so abundantly observed and described as
rotations.  Their energies may form rotational sequences, perhaps
somewhat modified, but corresponding transition probabilities do not
necessarily also follow the predictions of the rotational model.  This
is briefly discussed in the present investigation.  Closer inspection
of the transition probabilities between the expected rotational states
in heavier nuclei could reveal a similar behavior as for $^{8}$Be.
Such projects should be formulated and carried out, although we
anticipate this would be very difficult for these many-body systems.

A short-term direct perspective of the present investigation is
to apply the understanding to more complicated cluster structures like
three $\alpha$-particles.  This is straightforward with the
hyperspherical adiabatic expansion method, although numerically much
more elaborate. More generally, the lessons about transitions between
continuum states must be incorporated in analyses and interpretations
of the corresponding (few- and many-body) experimentally and
theoretically obtained structures.

Another short-term application is related to the extracted structure
information obtained from transition matrix elements.  We have
formulated a simple procedure to divide the contributions into four
pieces, that is the wave function is a sum of short-distance and
regularized asymptotic parts.  The tempting interpretation is
corresponding to resonance-resonance, continuum-resonance,
resonance-continuum, and continuum-continuum contributions.
Substantial contributions are found for the resonance-to-continuum
matrix element even when both initial and final state energies are
chosen to be precisely at the resonances.  The interference between
the various terms constitute a major contribution to the total
transition probability.

However, the interpretation of this division cannot be taken too far
for two reasons.  First, because far away from resonance energies the
short-distance contributions may still dominate, and hence not qualify
as a resonance contribution.  Second, the transition probability is
obtained by squaring the matrix element which necessarily further
entangles the division between resonance and continuum contributions.
We believe that appropriately defined energy windows combined with the
suggested division will be helpful in future analyses of experimental
data where continuum background contributions should be separated from that
of the pure resonance structure.  This perspective deserves much more
attention in future investigations.

In conclusion, rotational bands embedded in the continuum may still be
a meaningful concept but unexpected tendencies and significant
deviations from schematic model predictions can be present.  This
warning is so far only based on the decay and structure of the
$^{8}$Be two-body system.  The traditional rotational structure
investigations of the $^{8}$Be excited states has to be quantitatively
substantially modified.  The scarce experimental evidence supports the
present theoretical interpretation.  In general, the continuum
background plays an important role, and should be separated out in
analyses where only resonance properties enter.  On the other hand,
corresponding contributions can probably not be avoided, and has
therefore to be included.  \\

\acknowledgments This work was partly supported by funds provided by
DGI of MINECO (Spain) under contract No. FIS2011-23565.  We appreciate
valuable continuous discussions with Drs. H. Fynbo and K. Riisager.

\end{document}